\documentclass[12pt]{iopart}

\usepackage{graphicx}

\usepackage{todonotes,verbatim}

\usepackage{iopams}

\usepackage{hyperref}

\hypersetup{colorlinks=true,urlcolor=blue,linkcolor=blue,citecolor=blue}

\usepackage{url}
\usepackage{color}

\date{\today}
\begin{document}

\title{Negative plasmon dispersion in 2$H$-NbS$_2$ beyond charge-density-wave interpretation}

\author{Pierluigi Cudazzo$^{1,2}$, Eric M\"uller$^3$, Carsten Habenicht$^3$, Matteo Gatti$^{1,2,4}$, Helmuth Berger$^5$, Martin Knupfer$^3$, Angel Rubio$^{6,7}$, Simo Huotari$^8$}

\address{$1$ Laboratoire des Solides Irradi\'es, \'Ecole Polytechnique, CNRS, CEA,  Universit\'e Paris-Saclay, F-91128 Palaiseau, France}
\address{$2$ European Theoretical Spectroscopy Facility (ETSF)}
\address{$3$ IFW Dresden, P.O.Box 270116, D-01171 Dresden, Germany}
\address{$4$ Synchrotron SOLEIL, L'Orme des Merisiers, Saint-Aubin, BP 48, F-91192 Gif-sur-Yvette, France}
\address{$5$ \'Ecole Polytechnique F\'ed\'erale de Lausanne (EPFL), Institut de Physique des Nanostructures, CH-1015 Lausanne, Switzerland}
\address{$6$ Max Planck Institute for the Structure and Dynamics of Matter, Luruper Chaussee 149, 22761 Hamburg, Germany}
\address{$7$ Nano-Bio Spectroscopy Group and European Theoretical Spectroscopy Facility (ETSF), Universidad del Pa\'is Vasco CFM CSIC-UPV/EHU-MPC DIPC, 20018 San Sebasti\'an, Spain}
\address{$8$ Department of Physics, P.O.Box 64, FI-00014 University of Helsinki, Finland} 

\ead{angel.rubio@ehu.es}

\begin{abstract}
  We examine the experimental and theoretical electron-energy loss spectra in 2$H$-Cu$_{0.2}$NbS$_2$ and find that the 1 eV plasmon in this
  material does not exhibit the regular positive quadratic plasmon dispersion that would be expected for a normal broad-parabolic-band
  system. Instead we find a nearly non-dispersing plasmon in the momentum-transfer range $q<0.35$ \AA$^{-1}$. We argue that
  for a  stoichiometric pure 2$H$-NbS$_2$ the dispersion relation is expected to have a negative slope as is the case for other transition-metal
  dichalcogenides. The presence of Cu impurities, required to stabilize the crystal growth, tends to shift the negative plasmon
  dispersion into a positive one, but the doping level in the current system is small enough to result in a nearly-non-dispersing plasmon. We conclude that a
  negative-slope plasmon dispersion is not connected with the existence of a charge-density-wave order in transition metal
  dichalcogenides.
\end{abstract} 


\maketitle
\section{Introduction}

The charge density wave (CDW) is a broken symmetry state of metals
induced by electron-phonon or by electron-electron interactions. The
ground state is the coherent superposition of electron-hole pairs,
and, as the name implies, the corresponding charge density is not
uniform but displays a periodic spatial variation. Usually, the CDW
phase transition is accompanied by the appearance of an energy gap or
pseudo gap in the electronic band structure, phonon softening and
lattice distortions.  Layered transition-metal dichalcogenides
(TMD)\cite{wilson69} are prototypical materials displaying CDW
instability and for this reason they have attracted considerable
interest in the last years\cite{arpes}. However, although extensively
studied, a clear understanding of the physical mechanism responsible
for the appearance of CDW order in this class of materials is still
missing\cite{rossnagel11}. In particular, several explanations have
been put forward: from Fermi surface nestings\cite{wilson74} to van
Hove singularities (saddle points) in the density of states
(DOS)\cite{rice75}, until a recent theoretical work that pointed to
the role of electron-phonon coupling, ruling out a pure electronic
mechanism for the CDW instability\cite{johannes}.

In general, since a generic electronic instability induces a strong
enhancement of the charge-charge response function, this quantity
plays a crucial role in detecting CDW order not only from a
theoretical point of view but also experimentally. As a matter of fact, being
related to the inverse macroscopic dielectric function, it can be
directly measured in electron energy loss spectroscopy (EELS) and in
inelastic X-ray scattering (IXS) experiments.

In this context, J.~van Wezel and coworkers\cite{vanwezel11} have
established a direct link between CDW instability and collective
excitations (plasmons) that represent the poles of the charge-charge
response function. EELS measurements in three
prototypical TMDs belonging to the 2$H$ family (2$H$-TaS$_2$, 2$H$-TaSe$_2$
and 2$H$-NbSe$_2$)\cite{vanwezel11,schusterphd,schuster09} revealed
a dispersion relation that had a negative slope, i.e., the plasmon energy
was found to decrease with an increasing momentum transfer. In this case
the dispersion is said to be negative. This 
is in contrast with the case of a homogeneous
electron gas (the jellium model), 
where the plasmon has a positive parabolic dispersion.
On the basis of a macroscopic semiclassical Ginzburg-Landau model, it was 
suggested that the negative dispersion in TMD is the consequence of
the collective charge fluctuations associated with the CDW order
identifying a direct connection between CDW instability and plasmon
dispersion.\cite{vanwezel11} 
This picture was supported by the fact that for 2$H$-NbS$_2$,
the only system belonging to the 2$H$ family that does not display CDW
order, the available experimental results pointed out that 
the plasmon dispersion had a positive slope\cite{manzke81}.

Nevertheless, a negative plasmon dispersion has been found also in
several other materials that do not display CDW instability, for
example doped molecular crystals\cite{cudazzo11}. Even alkali
metals\cite{vomfelde89,aryasetiawan94,fleszar97,huotari09,huotari11,cazzaniga11,loa11,mao11} 
have been shown to exhibit several deviations from the jellium model including negative plasmon dispersion.
 Starting from this
observation, a different interpretation of the negative plasmon
dispersion in 2$H$-TMDs has been given on the basis of first principle
calculations \cite{faraggi12,cudazzo12,cudazzo14}. As shown in Ref.~\cite{cudazzo12}, the unusual
dispersion in this class of materials is due to the peculiar behavior
of intraband transitions that contribute to the 
plasmon excitation. If the negative plasmon dispersion is a pure band
effect, a similar behaviour should also be present  in NbS$_2$
contrary to the experimental observation\cite{manzke81}. 
However, it is
important to note that  stoichiometric pure NbS$_2$ is 
difficult to grow
since 2$H$-NbS2 is unstable in the growth
process\cite{schusterphd} unless metallic impurities are present. 
A possible 
positive dispersion could be an effect of impurities that 
are required to stabilize the system. Such an effect of 
metal impurities on the 
intraband plasmons in TMDs was shown by utilizing the rigid-band model
and time-dependent density functional theory (TDDFT) \cite{cudazzo12}.
Good agreement between TDDFT calculations and IXS spectra at high energy and large momentum transfer
was indeed recently found for  NbSe$_2$ and Cu$_{0.2}$NbS$_2$ \cite{cudazzo14_2}.

Motivated by these observations, we combined EELS experiments and
first-principle TDDFT calculations, and investigated the collective excitations
and their dispersion as a function of the momentum transfer in
2$H$-Cu$_{0.2}$NbS$_2$ with the aim to clarify the origin of the
negative plasmon dispersion in TMDs. From our measurements we did not find
a positive quadratic dispersion in this system, in contradiction
to what has been reported before. Earlier TDDFT-based work \cite{cudazzo12} 
predicted that the 
dispersion would be in fact negative if the concentration of Cu would be small
enough. Based on our new experimental findings, that match well 
the prediction of TDDFT, we argue that the negative plasmon dispersion is a
general property of 2$H$-TMDs related to the particular band structure
of these materials and that there is no need to invoke a coupling with
the CDW. 

\section{Methods}
\subsection{Experiment}
The Cu$_{0.2}$NbS$_2$ crystal 
was grown by vapor transport
using iodine as a transport agent.\cite{levy83}
The stoichiometry was verified by standardless energy dispersive 
spectrometry using a Jeol JXA-8600 electron probe microanalyzer.
The EELS measurements along $\Gamma M$ crystallographic direction
($\mathbf{q}~||~[1 0 0]$) were performed with a 172-keV spectrometer
described in Refs.\, \cite{roth14,fink89}. 
The momentum resolution was set to 0.04 \AA$^{-1}$ and the energy 
resolution to 85 meV. The sample temperature was controlled using a 
He flow cryostat, and measurements were done at $T=20$~K and
at room temperature. No particular difference between the results
at different temperatures was found, except for a blueshift
of the plasmon energy by $\sim$0.04 eV upon cooling to 20 K.
In the following, we concentrate on the data taken at 20 K.
The electron beam spot size on the sample was $\sim$0.5 mm.
The quasielastic zero-loss peak line's tail was subtracted 
from the measured spectra by fitting a 
suitable Pearson VII function to the data for energy transfer ($\omega$) between
0.1 and 0.3 eV.

The crystal used in the experiment 
was found to have a well defined stoichiometry of 
Cu$_{x}$NbS$_2$, with energy-dispersive x-ray spectroscopy 
revealing $x=$0.17--0.18 with the Cu atoms sitting in the interstitial region between the NbS$_2$ layers (hence for the remaining
of the article we indicate here the value $x \approx 0.2$).
The high quality of the crystal was verified by measuring the
electron diffraction pattern.

\subsection{Computational details}

The microscopic complex dielectric function 
$\epsilon=\epsilon_1 + i \epsilon_2$ is related to the
susceptibility $\chi$ (Ref.\, \cite{onida}) by the relation:
$\epsilon^{-1}=1+v\chi$ ($v$ being the Coulomb potential). In TDDFT
$\chi$ is the solution of the Dyson like equation:
$\chi=\chi_0+\chi_0(v+f_{xc})\chi$, where $\chi_0$ is the Kohn-Sham
(KS) susceptibility expressed in terms of KS eigenenergies and
eigenfunctions, while $f_{xc}$ is the exchange-correlation kernel, for
which we use the random phase approximation (RPA) ($f_{xc}=0$)\footnote{We note that exchange-correlation effects beyond RPA in this class of materials are not changing qualitatively the results\protect\cite{cudazzo12,cudazzo14_2}.}. The
Fourier components of both $\chi$ and $\epsilon$ are matrices in terms
of the reciprocal lattice vectors $\mathbf{G}$. The macroscopic
dielectric function $\epsilon_M$ is given by
$\epsilon_M(\mathbf{q},\omega)=1/\epsilon^{-1}_{\mathbf{G}\mathbf{G}}(\mathbf{q},\omega)$,
where $\mathbf{q}$ is inside the first Brillouin zone. The loss
function $L(\mathbf{q},\omega)$ measured in EELS experiments is
directly related to the imaginary part of the inverse macroscopic
dielectric function through the following equation:
$L(\mathbf{q},\omega)=-\mathrm{Im} \, \epsilon^{-1}_M(\mathbf{q},\omega)$.
 
In the present work the KS eigenenergies and eigenfunctions used to
determine $\chi_0$ have been evaluated in the local-density
approximation (LDA) implemented in a plane-wave-based
code\cite{gonze}. In our calculations we adopt the experimental
lattice parameters\cite{moncton}.
We use Troullier-Martins and Hartwigsen-Goedecker-Hutter
norm-conserving pseudopotentials \cite{troullier} (with an energy
cutoff of 120 Ry). In the calculation of $\chi_0$
(Ref. \cite{marini}), we used a $24\times 24\times 12$ grid of
$\mathbf{k}$ points and included 100 bands. The macroscopic dielectric
function has been obtained inverting a matrix of 300 $\mathbf{G}$
vectors (those parameters lead to converged results for the response
function in the range of energies and momentum studied in the present
work). Finally, the electron doping induced in  NbS$_2$ by the Cu
atoms has been simulated by shifting the Fermi level upward according
to the rigid-band model (in the present case this is equivalent to add 0.4 electrons per unit cell). An intrisic lifetime broadening of 0.05 eV was used in the calculation of the complex dielectric functions, and the loss functions obtained from the them were additionally convoluted with a Gaussian with a full width at half maximum of 0.085 eV to match the experimental energy resolution.

\section{Results and discussion}

\begin{figure}
\centering
\vspace{0.5cm}
\includegraphics[width=0.7\columnwidth]{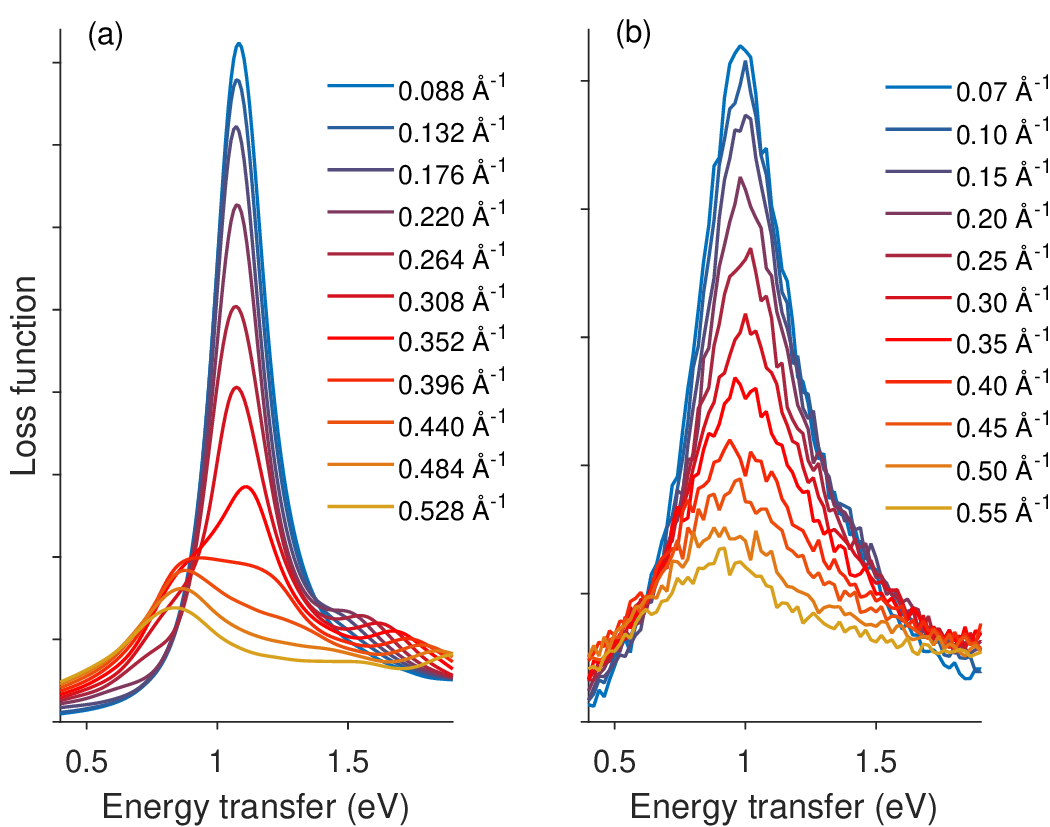}
\caption{(Color online) Loss functions for 2$H$-Cu$_{0.2}$NbS$_2$ based on (a) TDDFT calculation and (b) EELS experiment for several momentum transfers indicated in the legends. }
\end{figure}

The computed and measured loss function in 2$H$-Cu$_{0.2}$NbS$_2$ are
shown in Fig.~1 for various momentum transfers along the $\Gamma M$ direction.  
In
both cases, the loss spectrum is characterized by a peak that resides
at $\sim$1 eV. This structure is related to the collective excitations of the $d_z$ states of the Nb atoms\cite{cudazzo11}. It should be noted that the plasmon frequency obtained from the electronic density corresponding to these states (2.4 electrons per unit cell) overestimates the observed
frequency by about 4 eV. This overestimation is related to the presence of interband transitions that screen the bare plasmon frequency and are absent in the jellium model.
The peack is decreasing in intensity and broadening when the momentum transfer is increased. 
 The dispersion of its energy as a
function of momentum transfer is nearly negligible when $q<0.35$ \AA$^{-1}$. 
The results of the
experiment and theory are in excellent agreement.
Interestingly, the
dispersion seems to show a somewhat anomalous behavior 
for $q > 0.35$ \AA$^{-1}$. From the experimental spectra one might
conclude that the plasmon energy  starts to 
shift to smaller values with increasing $q$ above $q>0.35$ \AA$^{-1}$.
The reason for this becomes apparent when the theoretical spectra 
are carefully examined. At $q=0.352$ \AA$^{-1}$, the theoretical loss function 
peak position increases rapidly with an increasing $q$, with a
lower-energy peak appearing at $\sim$0.8--0.9 eV. This lower-energy
peak steals most of the spectral weight at the highest studied $q$. 
The experimental spectra show the same overall behavior, while due to 
the generally broader features the two structures are not 
individually resolved. However, comparison with the theoretical spectra
reveals uniquely the reason for the seemingly negative plasmon behavior
in this momentum-transfer range.

\begin{figure}
\vspace{0.5cm}
\centering
\includegraphics[width=0.7\columnwidth]{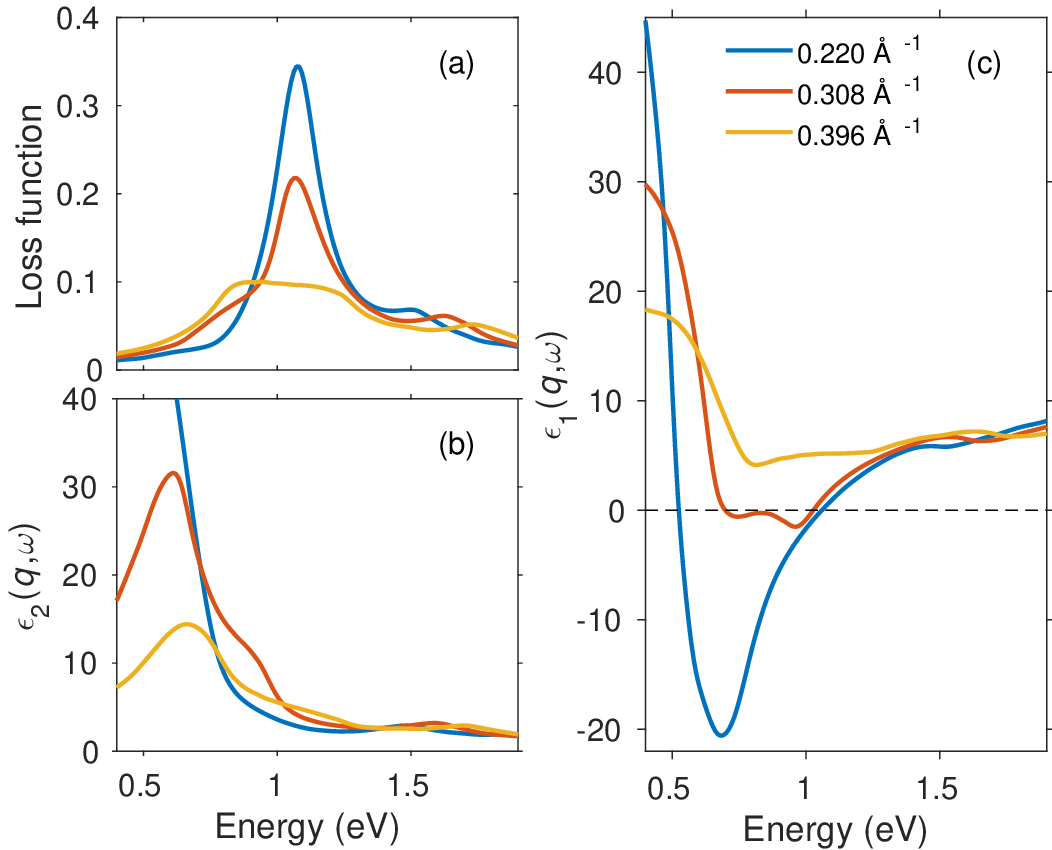}
\caption{(Color online) (a) The loss function, (b) the imaginary and (c) real part of the dielectric function for Cu$_{0.2}$NbS$_2$ for three selected momentum transfers.}
\end{figure}

To gain further insight into the behaviour of the plasmon in this
range of momentum transfer, we compare in Fig.~2 the loss function
with the real ($\epsilon_1$) and imaginary
($\epsilon_2$) parts of the dielectric function evaluated
between 0.22 and 0.4 \AA$^{-1}$.
As we can see, for a small  $q$, 
the 1-eV plasmon is caused by $\epsilon_1$ crossing zero.
This behavior has been shown previously to correspond to 
intraband transitions involving bands that cross the Fermi energy\cite{cudazzo11}. 
However, at $q=0.31$ \AA $^{-1}$ a shoulder appears in
$\epsilon_2$ just above the main intraband peak. Because $\epsilon_1$
and $\epsilon_2$ are related through the causality relation, this
shoulder causes an
oscillation in the real part of the dielectric function. This, in turn,
increases the number of zero crossings of the real part. 
This explains the 
appearance of a double structure in the loss function as seen in Fig.~1.
Thus, beyond $q$ = 0.35--0.4 \AA$^{-1}$  it is not
possible to clearly identify the position of the plasmon frequency
since in this range of momentum transfer the  loss function obtains
a relatively complex multiple-peak structure. At larger $q$ the
oscillation in $\epsilon_1$ disappears.
However, for
$q>0.40$ \AA$^{-1}$ the real part of the dielectric function
does no longer cross zero.  This means the plasmon is no
longer well defined since it decays into electron-hole excitations, and
the e-h continuum is probed.  It is noteworthy that even at smaller
momentum transfers the plasmon is strongly damped.

\begin{figure}
\centering
\vspace{0.5cm}
\includegraphics[width=0.7\columnwidth]{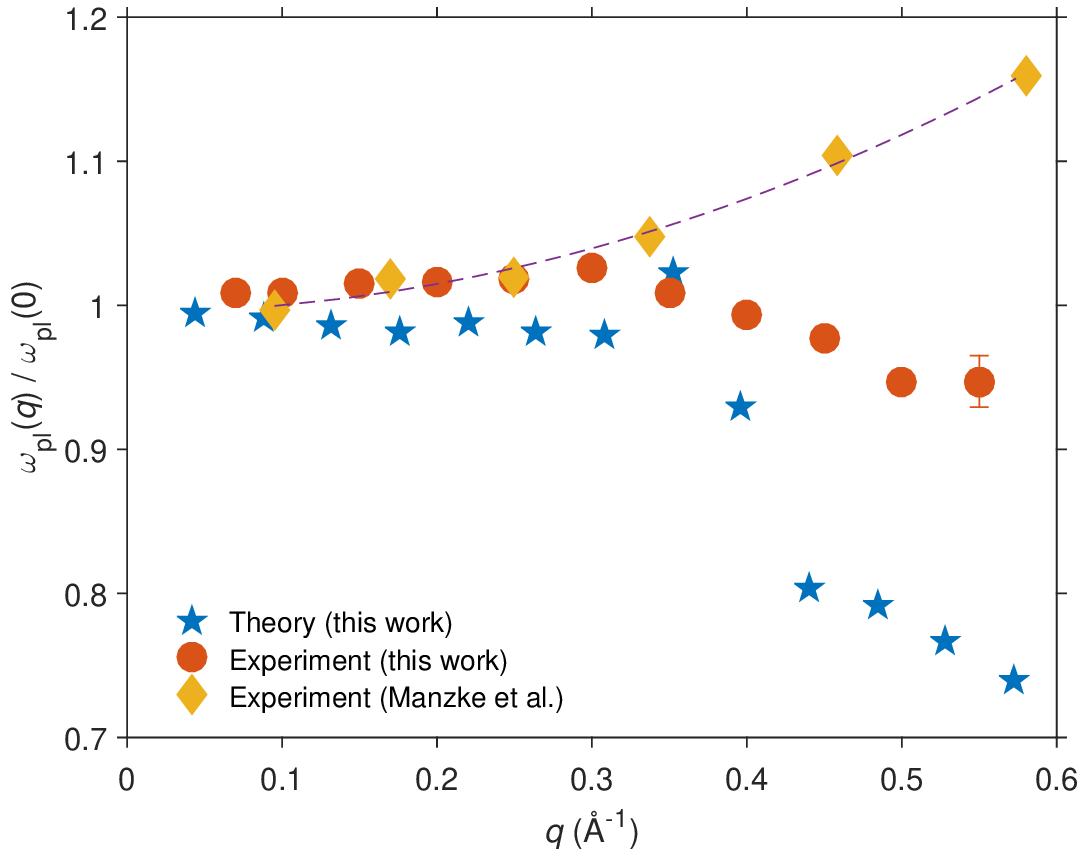}
\caption{(Color online) Plasmon dispersion for Cu$_{0.2}$NbS$_2$ 
as a function of $q$ for experiment,
theory, and compared to published results for NbS$_2$
(Ref.~\cite{manzke81}). The dashed line is a quadratic fit to the data points  of
Ref.~\cite{manzke81}.}
\end{figure}

The extraction of the dispersion from the plasmon is, 
based on the discussion above, a convoluted problem
owing to the peculiar behavior of the dielectric function and the
appearance of the doublet feature. We nevertheless present here the 
{\em apparent} energy of the loss-function peak as a function of $q$, 
of both experimental and theoretical
spectra. The peak position has been determined by fitting a Gaussian peak near the top of the plasmon peak in both cases. The uncertainty of the experimental peak positions has been determined by simulating how much the resulting peak position varies when additional random noise within the statistical errorbar is added to the data, performing the fit multiple times in this way, and taking the statistical 1$\sigma$ of the result as the uncertainty of the fit.
This dispersion is shown in Fig.~3, for plasmon energies
renormalized to values extrapolated to $q=0$, i.e., 
$\omega_\mathrm{pl}(q)/\omega_\mathrm{pl}(0)$,
which from the various sources is 
$\omega_\mathrm{pl}(0) \approx 0.95$ eV (Ref. \cite{manzke81}), 
$\approx 0.98$ eV (current experiment), and 
$\approx 1.09$ eV (current theory). 
From Fig. 3 it can be seen that while the published results from
Ref.~\cite{manzke81} show a positive, close to quadratic, 
upward slope of the plasmon
energy versus $q$ as expected for a plasmon in
homogeneous electron gas, our experiment and theory show much weaker
dispersion until $q=0.3$ \AA$^{-1}$, 
after which the dispersion becomes negative, owing to the 
reasons explained above.

It was theoretically shown by some of us \cite{cudazzo12} that 2$H$-TMDs in general 
are expected to exhibit negative plasmon dispersions for pure 
stoichiometric systems, but metallic dopants are expected to 
affect the dispersion curve, switching it to a positive dispersing
one. This is in agreement with experimental results on several doped TMD compounds (Refs.\, \cite{schusterphd,konig12,bu}). 
In our case, therefore, it can be deduced that in Cu$_{0.2}$NbS$_2$ the level of Cu-doping
is enough to flatten the dispersion curve, but not high enough 
to switch it into a monotonically increasing one.

The difference between our new results and 
than those obtained in Ref.\, \cite{manzke81} should be attributed
to the presence of impurities in the samples of both studies. 
However, in our case it can be argued that the low density of 
Cu impurities is not enough to fully switch the plasmon dispersion
into a quadratic one. This result is expected based on the prediction
of modeling the doping by a rigid-band shift as was done in 
Ref.\, \cite{cudazzo12}.
 
\section{Conclusions}
We have shown that in 2$H$-Cu$_{0.2}$NbS$_2$ the plasmon dispersion
curve is relatively flat instead of a positive dispersion observed in
a previous study. We have compared the experimental loss function of
2$H$-Cu$_{0.2}$NbS$_2$ to one calculated within the TDDFT, which we in
turn analyze in detail by breaking it up into the real and imaginary
part of the dielectric function. Together with comparison to earlier
literature on other TMDs of the 2$H$ family we conclude that our results corroborate the picture proposed by some of us in Ref.\cite{cudazzo12}: the measured negative plasmon dispersion is a basic property of all TMDs belonging to the 2$H$ family and is directly related to the shape of the bands crossing the Fermi level that is similar in this class of materials. Moreover, as it has been shown in previous theoretical and experimental works\cite{cudazzo12,schusterphd,konig12,bu}, electron doping beyond a given threshold, which depends on the system, switches the plasmon dispersion to have a positive slope. This means that an inclusion of metallic dopants, required to stabilize NbS$_2$ upon growth, will affect the dispersion curve, having a tendency to switch
it to a monotonically increasing one.
In conclusion, these results support the interpretation that the negative plasmon dispersion is a property of all undoped 2H compounds, independently of having a CDW order or not. Therefore it should not be attributed directly to a possible interplay of the electronic excitations and CDW ordering.

\ack

S.H. was supported by the Academy of Finland 
(projects 1254065, 1283136, 1259526 and 1295696), M.G. by a Marie Curie FP7 Integration Grant 
within the 7th European Union Framework Programme, P.C. by the European Union's Horizon 2020 
research and innovation programme under the Marie Sklodowska-Curie grant agreement No 660695. 
A.R. acknowledges financial support from the European Research Council
(ERC-2010-AdG-267374), Spanish grant (FIS2013-46159-C3-1-P), Grupos Consolidados
(IT578-13), and AFOSR Grant No. FA2386-15-1-0006 AOARD 144088, H2020-NMP-2014
project MOSTOPHOS, GA no. SEP-210187476 and COST Action MP1306 (EUSpec).
Computational time was granted by GENCI (Project No. 544).


\end{document}